\documentclass[aps,prstab,twocolumn,superscriptaddress,showkey]{revtex4}

\usepackage{graphicx}
\usepackage{amsmath}
\usepackage{bm}
\usepackage{epstopdf}

\begin{document}

\title{Characterization and Tuning of Ultra High Gradient Permanent Magnet Quadrupoles}

\author{S.~Becker}
\thanks{Corresponding author}
\email{stefan.becker@physik.uni-muenchen.de}
\affiliation{Ludwig-Maximilians-Universit\"at M\"unchen, 85748 Garching, Germany}
\author{M.~Bussmann}
\affiliation{Forschungszentrum Dresden-Rossendorf FZD, 01314 Dresden, Germany}
\author{S.~Raith}
\author{M.~Fuchs}
\author{R.~Weingartner}
\affiliation{Ludwig-Maximilians-Universit\"at M\"unchen, 85748 Garching, Germany}
\author{P.~Kunz}
\affiliation{Institut f. Kernphysik, Universit\"at Mainz, 55099 Mainz, Germany}
\author{W.~Lauth}
\affiliation{Institut f. Kernphysik, Universit\"at Mainz, 55099 Mainz, Germany}
\author{U.~Schramm}
\affiliation{Forschungszentrum Dresden-Rossendorf FZD, 01314 Dresden, Germany}
\author{M.~El Ghazaly}
\affiliation{Institut f. Kernphysik, Universit\"at Mainz, 55099 Mainz, Germany}
\author{F.~Gr\"uner}
\affiliation{Ludwig-Maximilians-Universit\"at M\"unchen, 85748 Garching, Germany}
\affiliation{Max-Planck-Institut f. Quantenoptik, 85748 Garching, Germany}
\author{H.~Backe}
\affiliation{Institut f. Kernphysik, Universit\"at Mainz, 55099 Mainz, Germany}
\author{D.~Habs}
\affiliation{Ludwig-Maximilians-Universit\"at M\"unchen, 85748 Garching, Germany}

\begin{abstract}
The application of quadrupole-devices with high field gradients and small apertures requires precise control over higher order multipole field components. We present a new scheme for performance control and tuning, which allows the illumination of most of the quadrupole-device aperture because of the reduction of higher order field components. Consequently, the size of the aperture can be minimized to match the beam size achieving field gradients of up to 500 $\rm{T\,m^{-1}}$ at good imaging quality. The characterization method based on a Hall probe measurement and a Fourier analysis was confirmed using the high quality electron beam at the Mainz Microtron MAMI.
\end{abstract}

\maketitle

\newcommand\degrees[1]{\ensuremath{#1^\circ}}

\newcommand{\comment}[1]{}

\section{Introduction}

High field gradient compact quadrupole-devices have recently been the subject of an increasing amount of attention, in particular, as a compact element for beam manipulation in laser based particle acceleration. Permanent-magnet quadrupole-devices (PMQs) with a small aperture can reach high magnetic field gradients and still maintain high surface magnetization. A number of design approaches have been developed and realized such as pure PMQs \cite{eichner,Lim} in accordance with a Halbach design \cite{halbach} or as modified (hybrid) Halbach quadrupole-devices utilizing saturated iron to guide the magnetic field \cite{Miha,Sears}.

While being of importance in compact accelerator setups, the main interest in PMQs lies in focusing particle beams of high divergence such as laser accelerated ion beams \cite{schollmeier} and electron beams \cite{nature1,nature2,nature3}. The control of the field quality as introduced in this work opens the path for using PMQs as focusing elements in Free-Electron-Lasers (FELs) \cite{ttfel} having a demand on high quality beam transport systems.
Multipole field components higher than the quadrupole field component have distorting effects on the electron beam and therefore increase the beam emittance. These higher order multipole field components (HOMFC) have to be minimized.

Assuming a constant ratio of the HOMFC and the pure quadrupole field component at a given radius, small aperture approaches typically suffer from a strong influence of the HOMFCs on the beam quality as the beam size to aperture is large compared to commonly used electromagnetic quadrupole-devices.

We present a method of tuning PMQs in order to achieve control over higher order field components, this allows the significant reduction of HOMFC and thus allows a large ratio of beam size to aperture.

\begin{figure}[ht]
\centerline{
  \centering
  \includegraphics[width=3cm]{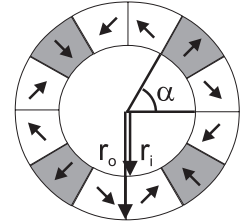}
}
\caption[Design of miniature permanent-magnet quadrupole-devices]{\label{design}Design of a miniature PMQ is shown with 12 wedges of permanent-magnet. The radius of the aperture is $r_{i}$ = 3 mm and the outer radius is $r_{o}$ = 10 mm. The arrows point in the magnetization direction.
}
\end{figure}

Halbach style PMQs where built using 12 wedges (Fig.~\ref{design}). The permanent-magnet material is NdFeB \cite{VAC} with a remanent field of 1.3 T. The assembled PMQ reaches surface magnetization fields of 1.5 T. The ability to reduce the aperture size compared to electromagnetic quadrupole-devices allows the realization of field gradients of up to 500 $\rm{T\,m^{-1}}$ at an aperture diameter of 6 mm. Conventional electromagnetic quadrupole-devices require a larger aperture and yield thus gradients of typically only 50 $\rm{T\,m^{-1}}$. These devices, however, usually allow the gradient to be adjusted which is not possible with a simple approach using PMQs. The PMQs as applied here were preliminarily tested and presented in \cite{eichner}.

Small apertures pose challenges in the measurement of the magnetic field distribution within. Common approaches involve the application of Hall probes to determine the field gradients or rotating coils to determine HOMFCs. This poses challenges in fabrication of the miniature coil and, in particular, suppressing vibrations during the measurement \cite{Datzmann,Iwa}.
We present a method allowing the measurement of all relevant magnetic vector field components relying solely on a miniature Hall probe which can be applied to very small apertures at the precision required.

The ability to measure all relevant field components within small apertures allows the introduction of specific HOMFCs by changing the position of individual magnet segments. We are thus able to compensate for undesired field components and also deliberately introduce specific components such as octupoles for compensating spherical aberrations or duodecapoles for compensating the effect of fringe fields. In order to minimize the influence of the correction of one field component on the entire field distribution, we apply materials with negligible non-linear interactions with the magnetic field due to hysteresis effects. Finally, we present measurement results of the tuning of the magnetic field distribution.

\section{Measuring Field Components}

The principle presented here for the measurement of the magnetic field involves a Hall probe with an active area of 200 $\mu m$ in diameter.

\begin{figure}[h]
\centerline{
  \centering
  \includegraphics[width=7cm]{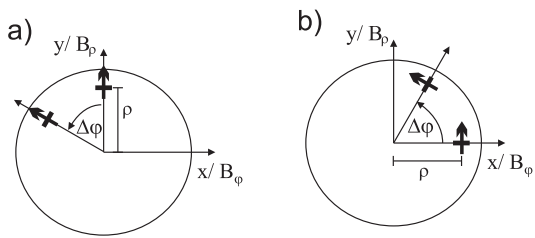}
}
\caption[Field measurement method]{\label{measurement}Measuring scheme of the magnetic vector field in cylindrical coordinates using a Hall probe with the arrow being the surface normal. The radial component (a) as well as the azimuthal component (b) are obtained separately.
}
\end{figure}

The magnetic field is scanned in cylindrical coordinates, as shown in Fig.~\ref{measurement}: The PMQ is mounted on a rotating stage for controlling the $\varphi$-coordinate. From the center of rotation, a displacement of the Hall probe along the y-axis scans the radial field component $B_{\rho}$, whereas the displacement along the x-axis scans the azimuthal component $B_{\varphi}$. The offset of $\varphi$ = \degrees{90} between $B_{\rho}$ and $B_{\varphi}$ has to be considered.

This method requires the knowledge of the position of the geometrical center of rotation which does not necessarily coincide with the center, i.e. the minimum value, of the magnetic field distribution.
In principle, a gradient of 500 $\rm{T\,m^{-1}}$ can be determined with sub-micrometer precision in spite of the active Hall probe area having a diameter of 200 $\rm{\mu m}$. The radius $\rho$ was chosen to be 1 mm, which was the maximum value for the measurements presented here due to the physical size of the specific Hall probe device applied. The tilt error of the probe limits the accuracy to sub 5 $\rm{\mu m}$ precision for the simple setup used here. The procedure for finding the geometrical center involves a simple feedback algorithm which only requires the Hall probe signal. The result of this iteration is unique as the field changes monotonously from an arbitrary point inside the aperture.

\subsection{Fourier Analysis}\label{secFourier}

A direct measurement of the entire magnetic field for $0<\rho<\rho_{0}$ in cylindrical coordinates inside the aperture over-determines the magnetic vector field. The assumption of $\rm{B_{z}}$ = 0 leads to the expansion of the magnetic field using polar coordinates of
\begin{equation}
\vec B(\rho,\varphi) = \sum_{l=1}^{\infty} [B_{l\rho}(\rho, \varphi)\vec{e}_{\rho}+ B_{l\varphi}(\rho, \varphi)\vec{e}_{\varphi}]
\label{vecField}
\end{equation}
with
\begin{align}
B_{l\rho}(\rho, \varphi) =&\;\rho^{l-1} [a_{l} \sin(l\varphi)+b_{l}\cos(l\varphi)]\\
B_{l\varphi}(\rho, \varphi) =&\;\rho^{l-1} [a_{l} \cos(l\varphi)-b_{l}\sin(l\varphi)],
\label{vecComps}
\end{align}
$a_{l}$ and $b_{l}$ being coefficients representing the HOMFCs. The case $\rm{B_{z}}\ne 0$ would imply fringe fields, which are discussed in the next section.

\begin{figure}[ht]
\centerline{
  \centering
  \includegraphics[width=7cm]{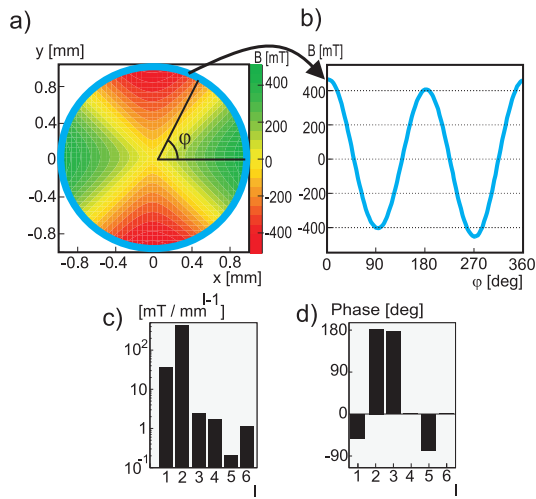}
}
\caption[Field measurement]{\label{ring}(a) shows field measurements using the azimuthal component in cylindrical coordinates and (b) its outer-most ring at $\rho$ = 1 mm is plotted against one rotation and used to expand the field coefficients $a_{l}$ and $b_{l}$, shown in (c) magnitude $\sqrt{a_{l}^{2}+b_{l}^{2}}$ and (d) phase $\arctan(b_{l}/a_{l})$.
}
\end{figure}

Measuring either the $\rm{B_{\rho}}$ or the $\rm{B_{\varphi}}$ field component on a single ring (Fig.~\ref{ring}a, b) is sufficient for a complete determination of the magnetic vector field. A Fourier expansion of a ring with the radius $\rho_{0}$ leads to the desired coefficients $a_{l}$ and $b_{l}$ in magnitude $\sqrt{a_{l}^{2}+b_{l}^{2}}$ (Fig.~\ref{ring}c) and phase $\arctan(b_{l}/a_{l})$ (Fig.~\ref{ring}d) allowing to construct the vector field (Eq. \ref{vecField}) using either
\begin{align}
a_{l} =& \frac{1}{\pi} \int\limits_{0}^{2\pi} \rho_{0}^{1-l} B_{\varphi}(\rho_{0}, \varphi) \cos(l\varphi) d{\varphi}\nonumber\\
b_{l} =& -\frac{1}{\pi} \int\limits_{0}^{2\pi} \rho_{0}^{1-l} B_{\varphi}(\rho_{0}, \varphi) \sin(l\varphi) d{\varphi}
\label{coeffPhi}
\end{align}
or
\begin{align}
a_{l} =& \frac{1}{\pi} \int\limits_{0}^{2\pi} \rho_{0}^{1-l} B_{\rho}(\rho_{0}, \varphi) \sin(l\varphi) d{\varphi}\nonumber\\
b_{l} =& \frac{1}{\pi} \int\limits_{0}^{2\pi} \rho_{0}^{1-l} B_{\rho}(\rho_{0}, \varphi) \cos(l\varphi) d{\varphi}.
\label{coeffRho}
\end{align}

\begin{figure}[ht]
\centerline{
  \centering
  \includegraphics[width=7cm]{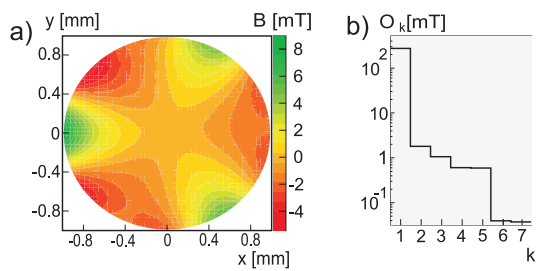}
}
\caption[Field measurement accuracy]{\label{accuracy} (a) shows the azimuthal HOMFC from Fig.~\ref{ring} ($\vec{B} = \sum_{l=3}^{6}[B_{l\rho}(\rho, \varphi)\vec{e}_{\rho}\,+\,B_{l\varphi}(\rho, \varphi)\vec{e}_{\varphi}]$, note the absolute scale compared to Fig.~\ref{ring}). (b) shows the remainder after the $\rm{k^{th}}$ field component.
}
\end{figure}

The two independent measurements of $\rm{B_{\varphi}}$ and $\rm{B_{\rho}}$ must yield the same values for the coefficients $a_{l}$ and $b_{l}$.
Knowing the absolute positioning error of the geometrical center, we obtain a relative error of $\Delta\rho/\rho = 0.5\%$ for $\rho_{0}$ = 1 mm.

The measurement errors for $\rm{B_{\varphi}}$ or $\rm{B_{\rho}}$ can be estimated by calculating the remainder

\begin{equation}
O_{k} = \int\limits_{0}^{2\pi} \left | B_{\rho/\varphi}(\rho_{0}, \varphi) - \sum_{l=1}^{k} B_{l\,\rho/\varphi}(\rho_{0}, \varphi) \right| d\varphi.
\label{eqRemainder}
\end{equation}

For all field measurements performed we find the contribution of orders above the sixth (the duodecapole) to be negligible as is shown by the remainder in Fig.~\ref{accuracy}b. This result is expected from the symmetry considerations of the design of the PMQ. Hence, the fact that measurement noise or signal drifts would significantly increase the remainder shows that those influences are clearly negligible up to at least the duodecapole order. A simple estimate of determining the maximum order component which can be resolved by a Hall probe with a diameter $d_{H}$ yields $l_{max} = \pi\rho_{0}/d_{H}$ = 15 and hence, a resolution well above the duodecapole ($l$ = 6).

\subsection{Fringe Fields}

The calculation of the field components from the ring measurement (Fig.~\ref{ring}) requires the assumption of $\rm{B_{z}}$ = 0 as mentioned before. Hence, the measurement of the ring used for the field expansion must not be performed in the fringes of the field distribution for the expansion following Eq. \ref{vecField}.

There are cases, however, where fringe fields cannot be neglected, in particular when particle beams are being focused to waist sizes on the nanometer scale. Fringe fields are discussed in \cite{halbach} for Halbach type PMQs. However, in practice, this does not determine the effect of fringe fields on a beam in a general way. Even if the design of the device and thus the field distribution including the fringe field is known in detail, the final effect on the beam still depends on the length of the PMQ and only works for specific beam properties, which in turn allows the determination of specific HOMFCs for compensating the effect of the fringe fields. The method presented here can be used to introduce field components in order to compensate the fringe field.

\section{Focus Measurement at MAMI Electron Beam}

The PMQs as introduced in \cite{eichner} have been applied at the accelerator MAMI to acquire the imaging quality. These measurements have been carried out prior to the ability to tune the devices. A PMQ-lens-doublet was used to focus the electron beam. The method of expanding the magnetic field distribution from a Hall probe measurement was applied to reproduce the experimental results.

The beam profile was monitored by a pair of 4 $\rm{\mu m}$ wires movable longitudinally in the direction of the beam propagation and transversely through the beam, both in the horizontal and the vertical direction. Bremsstrahlung caused by the beam hitting the wire was detected using an ionization chamber in the forward direction. The Bremsstrahlung's intensity, measured while changing the position of the wires transversely to the beam, determines the beam shape at a certain longitudinal position. The MAMI electron beam can reach energies of up to 855 MeV. We used energies of 270 MeV with an energy stability of $\Delta E/E = \rm{10^{-5}}$ and an emittance of 2 nm\thinspace rad horizontally and 0.7 nm\thinspace rad vertically.

The calculation of the beam transport involves an expansion of the magnetic fields of the quadrupoles following Eq. \ref{coeffPhi} and \ref{coeffRho} and tracking the electron beam \cite{GPT} using the the field map given by Eq. \ref{vecField}.

\begin{figure}[ht]
\centerline{
  \centering
  \includegraphics[width=8cm]{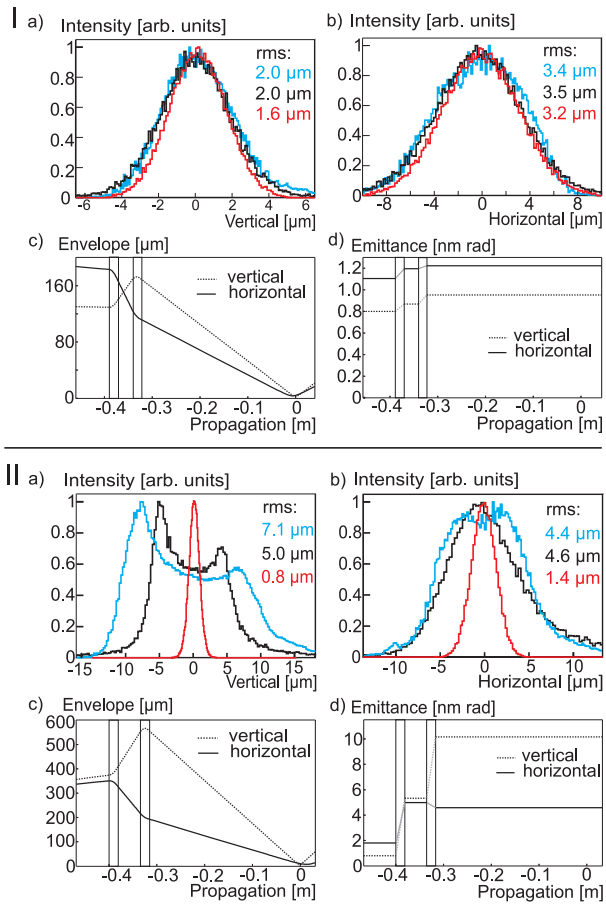}
}
\caption[MAMI Measurement]{\label{mami}Measurements at the Mainz Microtron MAMI are shown with two different beam configurations: The small electron beam configuration (panel I) is shown with the horizontal (Ia) and vertical (Ib) beam plane at the waist. The measured beam (blue), the calculated beam using expanded fields (black) and calculated beam using ideal fields (red) are shown for comparison. The calculated beam envelope (Ic) and the emittance (Id) are plotted against the propagation direction of the beam. The vertical lines mark the PMQ positions. The large beam configuration is shown correspondingly (panel II).
}
\end{figure}

Two beam configurations were chosen: A convergent electron beam of small size at the entrance of the lens-doublet (Fig.~\ref{mami}I), and a divergent beam of larger size (Fig.~\ref{mami}II). The measured waist size is only a little larger than that of an ideal quadrupole with the same gradient as expected for the small-beam configuration. The waist size of the transport calculation using the expanded fields agrees well with the measured beam waist. The emittance remains virtually constant.

Higher order field components at the large beam configuration significantly distort the electron beam. The ideal waist size is almost an order of magnitude smaller than measured (Fig.~\ref{mami}IIa,b). The effect of the HOMFC can be seen in the evolution of the trace space emittance \cite{Floettmann} in particular, as is shown in Fig.~\ref{mami}Id,IId. The emittance was deduced from the calculation using the expanded field distribution of the PMQ.

A quadrupole-lens-doublet imaging on the $\mu m$ scale is very sensitive to the quadrupole phases not matching with the horizontal/vertical beam plane of the accelerator. An offset in the order of 100 $\mu$rad already has a notable impact leading to an enlargement of the size of the beam waist. Hence, the major reason for the deviation between the experiment and the calculation of Fig.~\ref{mami}IIa is most likely due to the quadrupole phases being slightly off compared to the accelerator beam planes. Nevertheless, the experiment and the calculations show good agreement which confirms the field measurement method.

\section{Magnetic Field Tuning}

HOMFCs can have a variety of origins, for example variations of the shape of the wedges or the magnetization direction or strength. The knowledge of the specific origin of an undesired higher order field component is not necessary for its compensation: The introduction of a field component in the same order and magnitude but with a phase shift of \degrees{180} leads to its elimination.

Displacing certain wedges introduces well defined higher order field components which can be used for correcting manufacturing deviations of the wedges, the housing or for modeling the magnetic field distribution correcting imaging aberrations.

\subsection{Wedge Positioning}

\begin{figure}[ht]
\centerline{
  \centering
  \includegraphics[width=8cm]{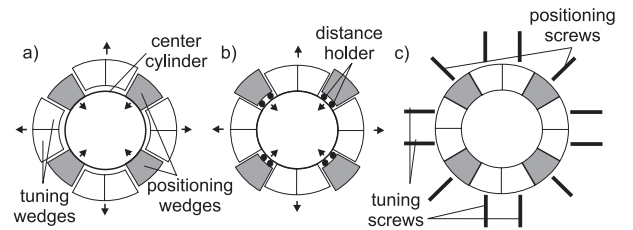}
}
\caption[Tuning method]{\label{schema}(a) shows an arrangement of permanent-magnet wedges and the center cylinder are shown schematically. The arrows point in the direction of the magnetic forces which act centrifugally and centripetally on the wedges. (b) Example for distance holders (e.g. 50-100 $\mu m$ aluminium foil) of the positioning wedges from the cylinder center. (c) Positioning and tuning screws acting on the wedges.
}
\end{figure}

The four wedges with the magnetic field oriented towards the device axis, which in the following are called positioning wedges, experience centripetal magnetic forces. A thin non-magnetic cylinder is placed inside the aperture. Its radius determines the radial distance of the wedges from the axis of the device (Fig.~\ref{schema}a,b). The positioning wedges are tightened with positioning screws from the housing for fixing the cylinder. The center of the magnetic field can be adjusted to coincide with the geometrical center of the PMQ for the elimination of the dipole field components.

\begin{figure*}[ht]
  \centering
  \includegraphics[width=17cm]{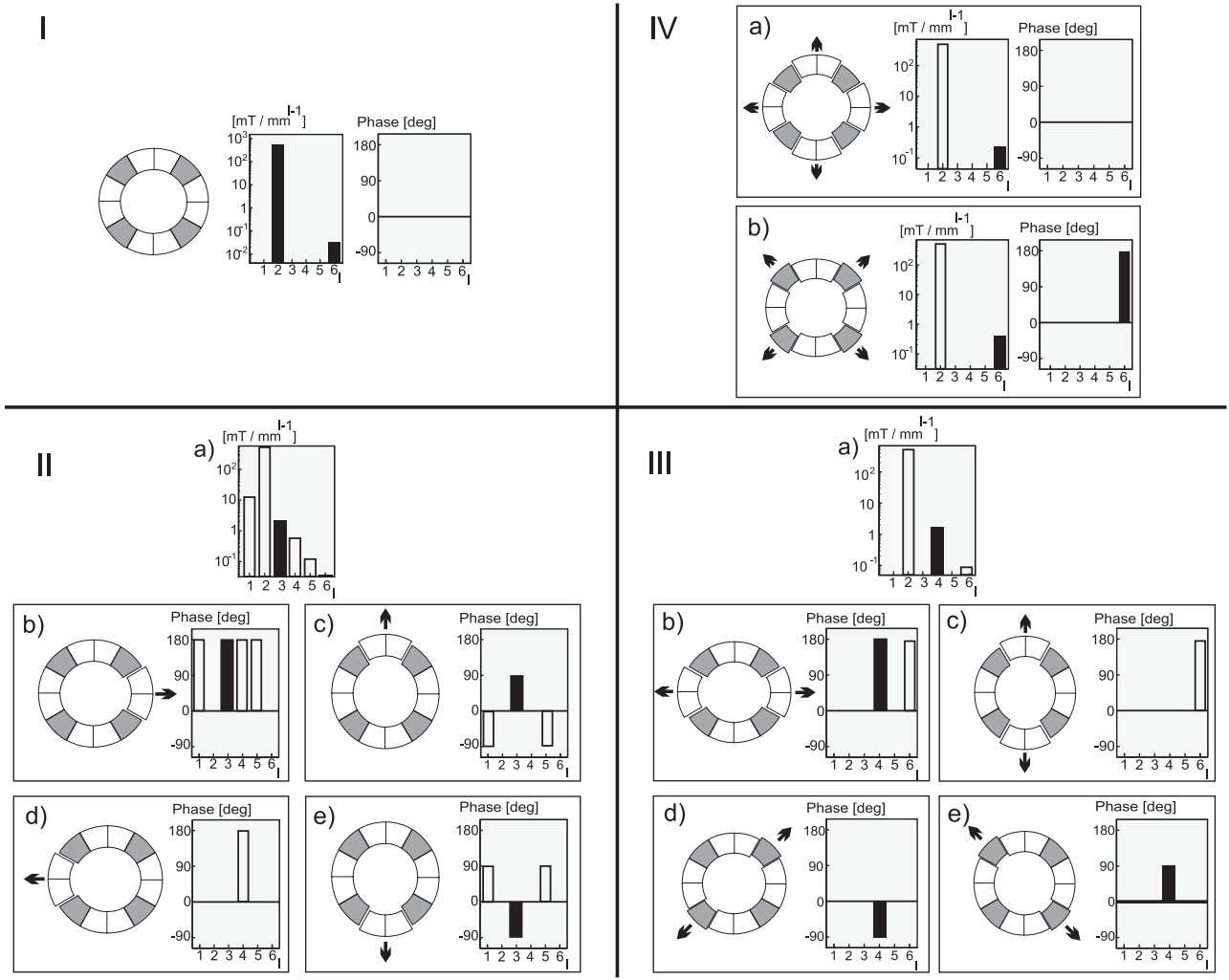}
\caption[Introduction of Higher-Order Field Components]{\label{adHOMFC}The introduction of HOMFC with the PMQ having the same orientation as in Fig.~\ref{design}. Panel I: Magnitude and phases of the calculated magnetic field using the ideal arrangement of permanent-magnet wedges. Panel II: Displacing a single pair of tuning wedges by 150 $\rm{\mu m}$ introduces a dominant sextupole (IIa). Independent from the specific pairs of tuning wedges displaced, panels IIb-IIe show the effect on the phases of the introduced field components by moving tuning wedges at $\alpha$ = \degrees{0}, \degrees{90}, \degrees{180}, \degrees{270}. Panel III: Two opposite pairs of tuning wedges are displaced by 150 $\rm{\mu m}$, this introduces a dominant octupole. Panel IIIa shows the effect on the magnitude of the field components. Panel IIIb-IIIc show the effect on the phases of the introduced field components moving tuning wedges at $\alpha$ = \degrees{0} and \degrees{90} affecting $a_{4}$. Moving the positioning wedges as shown in panels IIId-IIIe affect $b_{4}$. Panel IV: All pairs of tuning wedges are displaced by 150 $\rm{\mu m}$, this introduces a dominant duodecapole. The effect on the magnitudes are shown in panel IVa. Panels IVb and IVc show distinct pairs of tuning wedges which are moved and the effect on magnitudes and phases of the field components. The application of distance holders might be required as in Fig.~\ref{schema}b.}
\end{figure*}

Magnetic forces centrifugally repel the four remaining pairs of wedges, called tuning wedges in the following. In combination with tuning screws, these forces allow their precise positioning. Since the magnetic forces are acting on the tuning wedges as a pair, the tuning screws are arranged in parallel (Fig.~\ref{schema}c).

\subsection{Introducing of Field Components}

The field distribution is altered by modifying the PMQ by selecting a pair of tuning wedges and moving these. The field distribution of a modified PMQ is calculated numerically \cite{EMS} for obtaining the quantitative effect on individual field components including their phase. The result of the calculation is expanded (Eq. \ref{coeffPhi} and \ref{coeffRho}) and used for obtaining a table of reference for the effect of moving tuning pairs on the field distribution.

We first consider the undisturbed quadrupole design in Fig.~\ref{adHOMFC}I. Owing to the symmetry of the design, only a duodecapole superimposes on the quadrupole field. Fig.~\ref{adHOMFC}II shows a dominant sextupole ($l$ = 3) which is introduced when moving one pair of tuning wedges. For symmetry reasons, a pure sextupole component cannot be introduced, but an octupole component is also obtained which in turn can be eliminated. The introduction of an octupole component is achieved by moving opposite pairs of wedges as shown in Fig.~\ref{adHOMFC}III. Changing $b_{4}$ (Eq. \ref{vecComps}) without influencing $a_{4}$ requires the movement of the positioning wedges which can be achieved by e.g. introducing distance holders as is shown in Fig.~\ref{schema}. Alternatively, $b_{4}$ can be modified by moving two individual tuning wedges which are arranged at opposite locations from the device center which requires a subsequent compensation of the additionally introduced $a_{4}$ component. The introduction of a duodecapole field is shown in Fig.~\ref{adHOMFC}IV. Depending on the desired phase, the application of distance holders might be required as is shown in Fig.~\ref{schema}b.

\subsection{Adjustment Results}

\begin{figure}[ht]
\centerline{
  \centering
  \includegraphics[width=7cm]{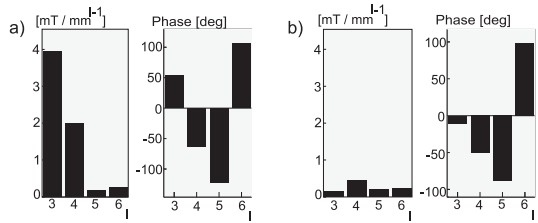}
}
\caption[Magnetic field component spectrum as a tuning result]{\label{tuning}(a) shows the HOMFC within a newly assembled quadrupole and (b) shows the HOMFC compensated.
}
\end{figure}

Fig.~\ref{tuning}a shows an example of a newly assembled quadrupole. Due to manufacturing deviations of either the wedges or the housing, there is a considerable initial sextupole field component.
After compensating for the higher order field components, a much purer quadrupole field is obtained as is shown in Fig.~\ref{tuning}, demonstrating the feasibility of the method discussed. The errors of these measurements correspond to those errors discussed in Fig.~\ref{accuracy}b, since the apparatus used is the same.

\section{Conclusion}

The method presented allows to shape the magnetic field distribution of a PMQ. The magnetic field distribution is determined from a Fourier expansion of a Hall probe measurement and used as the basis for identifying the tuning wedges to be moved for obtaining the desired field distribution. The precise quantification of HOMFC in conjunction with the complete control of the field configuration allows to accurately configure the magnetic field distribution to a high degree. After only a few iterations, magnitude and phase of the undesired field components can be reduced significantly. Hence, the control over these field components up to at least the duodecapole allows a larger ratio of the quadrupole's aperture to be illuminated. Moreover, HOMFCs such as an octupole or duodecapole can be introduced in order to compensate for imaging aberrations and fringe fields effects.

The advantage of the pure permanent-magnet devices over hybrid quadrupole designs lies in the linear superposition of the magnetic field contributions of the individual segments. This allows a decoupled tuning process, and thus a fast and simple adjustment of the magnetic field distribution. The compensation scheme shown here still has potential for improvement since the results presented in this publication were obtained by manually tuning the PMQs. The method for the reduction of HOMFCs can easily be automated using simple algorithms which allow to move the wedges at higher precision.

\section*{Acknowledgments}

This work has been funded by the DFG through transregio TR18 and supported by the DFG cluster-of-excellence Munich Center for Advanced Photonics MAP.

\end{document}